\documentclass[journal,twoside,web]{ieeecolor}
\usepackage{generic}
\usepackage{cite}
\usepackage{amsmath,amssymb,amsfonts}
\usepackage{algorithmic}
\usepackage{graphicx}
\usepackage{multirow}
\usepackage{threeparttable}
\usepackage{algorithm,algorithmic}
\usepackage{makecell}
\usepackage{hyperref}
\usepackage{textcomp}
\def\BibTeX{{\rm B\kern-.05em{\sc i\kern-.025em b}\kern-.08em
    T\kern-.1667em\lower.7ex\hbox{E}\kern-.125emX}}
\begin{document}
\title{Morphological-consistent Diffusion Network for Ultrasound Coronal Image Enhancement}
\author{Yihao Zhou, Zixun Huang, Timothy Tin-Yan Lee, Chonglin Wu, Kelly Ka-Lee Lai, De Yang, Alec Lik-hang Hung, Jack Chun-Yiu Cheng, Tsz-Ping Lam, Yong-ping Zheng
 \IEEEmembership{Senior Member, IEEE}
\thanks{
    This work involved human subjects or animals in its research.
    Human subject ethical approval was granted from local institutional review board (Joint Chinese University of Hong Kong - New Territories East Cluster Clinical Research Ethics Committee, CREC Ref. No. 2015.463 \& 2016.658 and The Hong Kong Polytechnic University, Ref. No. HSEARS20180906005-01).
}
\thanks{This study was partially supported by The Research Grant Council of Hong Kong (R5017-18).}
\thanks{Yihao Zhou, Timothy Tin-Yan Lee, Kelly Ka-Lee Lai,
    Chonglin Wu, De Yang, and Yong-Ping Zheng are with the Department of Biomedical Engineering, The Hong Kong Polytechnic University, Hong Kong, China. Yong-Ping Zheng is also with the Research Institute for Smart Ageing, The Hong Kong Polytechnic University, Hong Kong, China. (e-mail:\{yihao.zhou, timothy.lee\}@connect.polyu.hk; \{kelly.lai, chonglin.wu, de-derek.yang, yongping.zheng\}@polyu.edu.hk).}
\thanks{Zixun Huang is with the School of Artificial Intelligence, Shenzhen Polytechnic University, Shenzhen, China. (e-mail: zixunhuang@szpu.edu.cn)}
\thanks{Alec Lik-hang Hung, Jack Chun-Yiu Cheng, and Tsz-Ping Lam are with the SH Ho Scoliosis Research Lab, Joint Scoliosis Research Center of the Chinese University of Hong Kong and Nanjing University, Department of Orthopaedics and Traumatology, The Chinese University of Hong Kong, Hong Kong SAR, China. (e-mail: \{AlecLHhung, jackcheng, tplam\}@cuhk.edu.hk).}
}

\maketitle

\begin{abstract}
Ultrasound curve angle (UCA) measurement provides a radiation-free and reliable evaluation for scoliosis based on ultrasound imaging. However, degraded image quality, especially in difficult-to-image patients, can prevent clinical experts from making confident measurements, even leading to misdiagnosis. In this paper, we propose a multi-stage image enhancement framework that models high-quality image distribution via a diffusion-based model. Specifically, we integrate the underlying morphological information from images taken at different depths of the 3D volume to calibrate the reverse process toward high-quality and high-fidelity image generation. 
This is achieved through a fusion operation with a learnable tuner module that learns the multi-to-one mapping from multi-depth to high-quality images. Moreover, the separate learning of the high-quality image distribution and the spinal features guarantees the preservation of consistent spinal pose descriptions in the generated images, which is crucial in evaluating spinal deformities. Remarkably, our proposed enhancement algorithm significantly outperforms other enhancement-based methods on ultrasound images in terms of image quality. Ultimately, we conduct the intra-rater and inter-rater measurements of UCA and higher ICC (0.91 and 0.89 for thoracic and lumbar angles) on enhanced images, indicating our method facilitates the measurement of ultrasound curve angles and offers promising prospects for automated scoliosis diagnosis.
\end{abstract}


\section{Introduction}
\label{sec:introduction}

\IEEEPARstart{A}{dolescent} Idiopathic Scoliosis (AIS)  is a complex three-dimensional spinal deformity that predominantly affects adolescents aged 10-16 years during their growth phase. AIS has a global prevalence ranging from 0.47\% to 5.2\% \cite{konieczny2013epidemiology}. The current gold standard for diagnosing AIS involves measuring the Cobb angle through X-ray imaging \cite{cobb1948}. However, the ionizing radiation from X-rays poses health risks. Patients typically undergo at least 25 X-ray examinations during their treatment period \cite{oakley2019scoliosis}. As an alternative, 3D ultrasound imaging emerges as a promising non-radiative technique and has demonstrated reliability and feasibility for frequent and less hazardous evaluations of scoliosis progression \cite{lai2021validation, zheng2016reliability}. 
\begin{figure}
    \centering
\includegraphics[width=0.38 \textwidth,height=0.48 \textwidth]{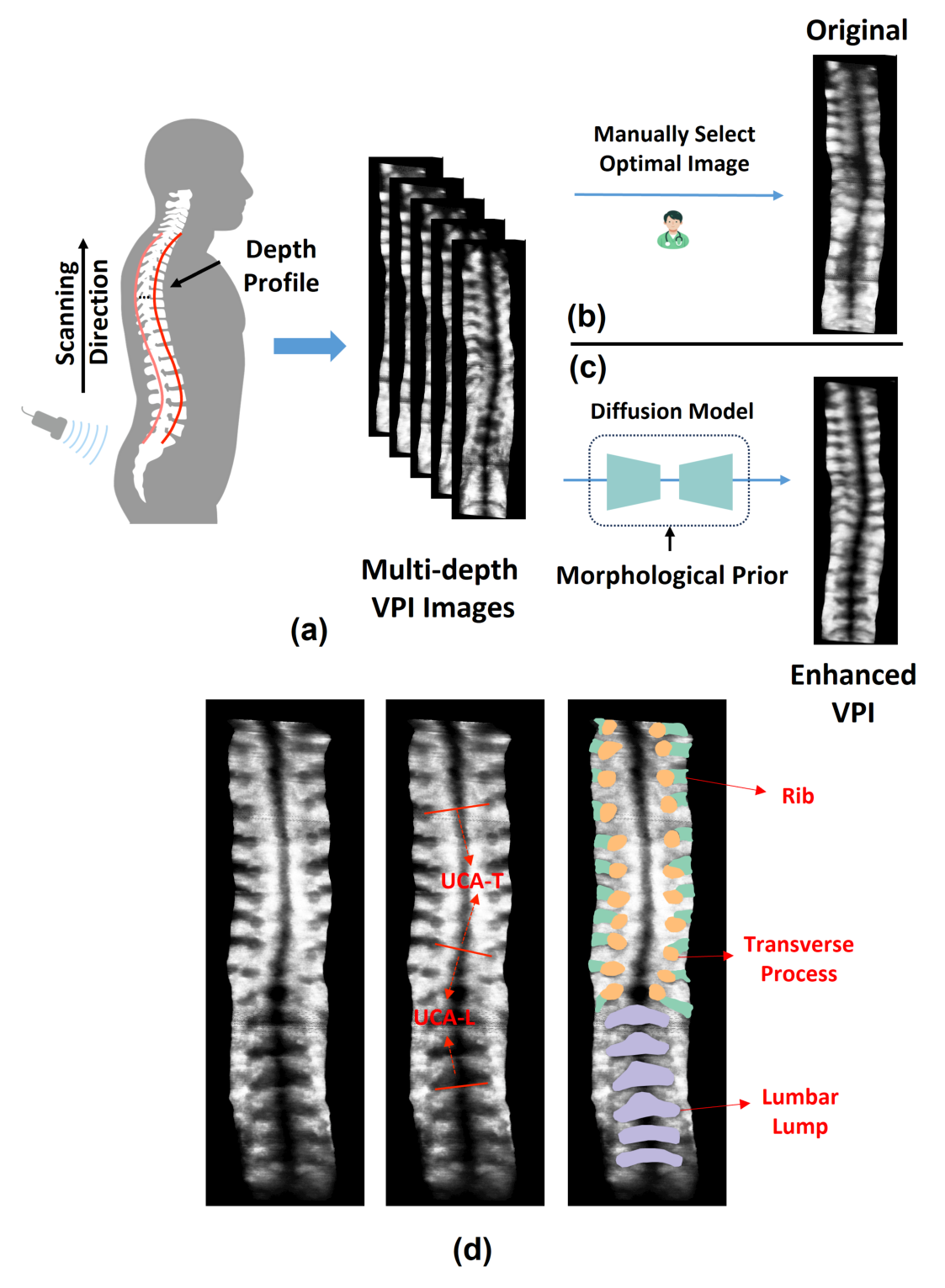}
    \caption{
    (a) Generating 2D projection images in the order of increasing depth (posterior to anterior) using a customized depth profile. 
    (b) Traditional workflow of scoliosis diagnosis using ultrasound image.
    (c) Our proposed image enhancement pipeline to improve the visual perception of the spine for reliable scoliosis diagnosis. 
    (d) Ultrasound Curve Angle (UCA) measurement and the segmentation of the spine. The lines are delineated on the most tilted region of the lumbar vertebrae shadow (UCA-L) and the paired thoracic transverse process (UCA-T). }
    \label{introduction}
\end{figure}

\begin{figure}
    \centering
\includegraphics[width=0.49\textwidth,height=0.31\textwidth]{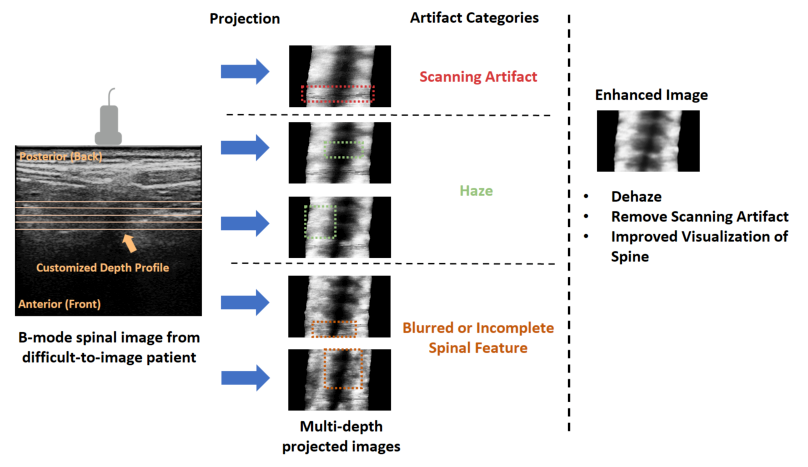}
    \caption{
        Common artifacts in the projected coronal ultrasound images from the difficult-to-image patient. The enhanced image provides improved visual perception of the spine and removes artifacts. 
    }
    \label{introduction_noise}
\end{figure}

In clinical practice, diagnosing AIS with 3D ultrasound can be achieved by mapping the 3D ultrasound volume onto a series of 2D coronal images based on various thicknesses and depths and performing measurements on the 2D coronal image with the best quality (Fig. \ref{introduction}(a)). This technique is referred to as Volume Projection Imaging (VPI) \cite{cheung2015ultrasound}.
Following the ultrasound curve angle measurement (UCA) protocol in \cite{lee20213d}, experts are required to select the optimal depth image and then identify the most tilted vertebrae (Fig. \ref{introduction}(b)). 
To streamline the measurement process, spine segmentation can serve as a valuable aid in aligning the placement of measurement lines \cite{xie2023structure} (Fig. \ref{introduction}(d)).
However, the quality of the acquired VPI images can be affected by many factors, such as speckle noise, the body mass index (BMI) of the subject, the selection of VPI images with different depths and fitness, etc., which prevents a single VPI image from guaranteeing that all the spinal features are visible, as shown in Fig \ref{introduction_noise}. These objective factors of image quality and subjective factors behind personal experience are inevitable in scoliosis diagnosis, resulting in large inter-observer and intra-observer variations. Furthermore, acoustic energy must pass through the skin, subcutaneous fat, multifidus muscle, and erector spinae muscle layers to reach the spine. The multipath reflections generated by these layers cause the spinal features to not always be visible in ultrasound images. This effect is particularly pronounced in patients who are technically difficult to image, such as those with obesity or dense muscular structures, often resulting in inconclusive examinations.
Therefore, UCA measurement based on VPI images typically demands extensive experience from the operators. Accordingly, the current clinical workflow can greatly benefit from an image enhancement algorithm that ensures the high quality and visibility of spinal features (Fig. \ref{introduction}(c)). 

One potential way to generate a high-quality VPI image from a 3D ultrasound volume is to analyze the entire 3D volume and project the voxels with spinal features onto the coronal plane. However, the high computational complexity and the requirement for dense labeling render it impractical in real-time diagnosis \cite{huang2022joint}. 
Image enhancement with a fusion of 2D VPI images at different depths is a promising and practical solution owing to its versatility and efficiency.
With the rapid development of deep learning techniques, Deep Convolutional Neural Networks (DCNNs) have shown their superiority in medical image enhancement. However, the enhancement based on DCNNs generally requires a large amount of paired data for learning. The image quality of VPI images is related to various factors, such as the experience of scanning operators, probing, VPI settings, etc. In practice, the acquired VPI images cannot produce the paired high-quality and low-quality reference of each observation. These limitations trigger us to investigate a weakly supervised image enhancement algorithm, especially for spinal feature enhancement. 

Generative models have been widely studied in weakly supervised image enhancement as they do not rely on paired training data and have achieved great success in MRI, CT, and ultrasound \cite{dissanayake2022generalized, zhou2019ultrafast, das2021sgan, khor2022ultrasound}. Traditional generative models, such as generative adversarial networks (GANs) and variational autoencoders (VAEs), aim to model the training data distribution. These enhanced algorithms depend highly on the training data and may struggle with robustness in real applications. As an alternative, diffusion models (DM) are based on a process that gradually transforms a simple noise distribution into a complex data distribution through a series of iterative steps. Diffusion models are trained using a well-defined probabilistic framework, ensuring stability and convergence. This makes them more reliable and easier to train than GANs, which require careful tuning to achieve stability. The iterative nature of diffusion models allows for the stable generation of high-quality and high-fidelity images, which is particularly beneficial for medical image enhancement, where capturing fine details and ensuring image fidelity are crucial.

In this paper, we conduct an in-depth study on diffusion-based VPI image enhancement and propose a novel weakly supervised image enhancement method without requiring paired high- and low-quality image pairs for training. As mentioned, the examiner will focus on the spinal features when utilizing ultrasound VPI images to assess spinal deformity. Different from traditional DM methods that utilize the low-quality image as the condition, we proposed to employ a segmentation model to extract the spinal features from the low-quality images and utilize the intermediate features from the segmentation model to facilitate image enhancement with an awareness of structural spinal features.  

To fully exploit the morphological information of the 3D ultrasound volume, we presented a multi-to-one image mapping strategy to incorporate all spinal-related features from images extracted from various depths. To fuse the intermediate spinal features from VPI images to the diffusion-based enhancement model for task-oriented image enhancement, we further proposed a morphology tuner module that is integrated within the skip connections of the enhancement network to enable precise control over the generated contours of anatomical structures. The main contributions of this paper are summarized as follows: 



\begin{itemize}
    \item 
    We proposed a structure-aware diffusion-based generative model for ultrasound VPI image enhancement. A spinal feature extractor was proposed to capture morphological features crucial in scoliosis diagnosis and calibrate the generative process toward high-quality and high-fidelity image synthesis. 
    \item 
    We presented a morphology tuner module integrated within the skip connections of the enhancement network, allowing for precise control over the generated contours of anatomical structures, specifically targeting task-oriented image enhancement.
    \item 
    We proposed a multi-stage training framework with a multi-to-one image mapping strategy to incorporate spinal-related features from various depths of 3D ultrasound volumes, maximizing the use of morphological information and preserving consistent spine anatomy. 
    
    \item We conducted extensive experiments to demonstrate that our proposed image enhancement strategy highlights the morphology details of spinal structure and thus improves the reliability of UCA measurement.
\end{itemize}

\section{Related works}

\subsection{Ultrasound Volume Projection Imaging for Scoliosis Assessment}
In scoliosis examinations, using a 3D ultrasound probe to capture the entire spine in a single shot provides crucial spatial information that is otherwise unavailable in 2D ultrasound, i.e. B-mode image. To visualize the whole spine anatomy, volume projection imaging was proposed to project the voxels of the 3D ultrasound volume onto a series of 2D coronal images via non-planar volume rendering \cite{cheung2015ultrasound}. However, using the spinal process for the measurement underestimated the deformity due to the axial rotation of the spine \cite{zheng2016reliability}. As the first attempt to assess scoliosis using 3D ultrasound, Chen et al. \cite{chen2013reliability} proposed manually measuring the curve of the spine profile in coronal images using the ultrasound reflection features of the lamina. However, not all the lamina features can be visible in the coronal images formed in this method due to the titling of the lamina surface and the attenuation of ultrasound when passing thick, soft tissues. As a promising alternative, ultrasound curve angle measurement is proposed to measure the spine deformity according to the paired thoracic processes and lumbar vertebrae \cite{lee20213d}. Recently, owing to the superiority of deeply learned features, many deep learning-based algorithms were proposed to facilitate the VPI-based scoliosis measurement. Lyu et al. \cite{lyu2021dual} presented a dual-task framework with boundary detection as an auxiliary task to regularize spine segmentation. Huang et al. \cite{huang2022joint} proposed a multi-task framework for joint noise removal and spine segmentation in VPI images. A semantic segmentation algorithm might make it difficult to measure the scoliosis degree automatically due to the lack of information, such as the bone index and spine vertebrae landmark coordinates. To achieve a fully automatic measurement of scoliosis, Zhou et al. \cite{zhou2024automatic} proposed a spinal landmark localization algorithm integrating affinity clustering for fully automatic spine deformity measurement. However, the aforementioned researches are all based on high-quality VPI image datasets. In a real application, the image quality of the obtained VPI images may be degraded, especially from difficult-to-image subjects. To ensure the robustness of deep learning-based approaches that are trained by high-quality image datasets, it is vital to develop an image enhancement algorithm that can guarantee the VPI image quality is comparable to that of the high-quality image dataset.  

\begin{figure*}[ht]
    \centering
    \includegraphics[width=0.8\textwidth,height=0.3\textwidth]{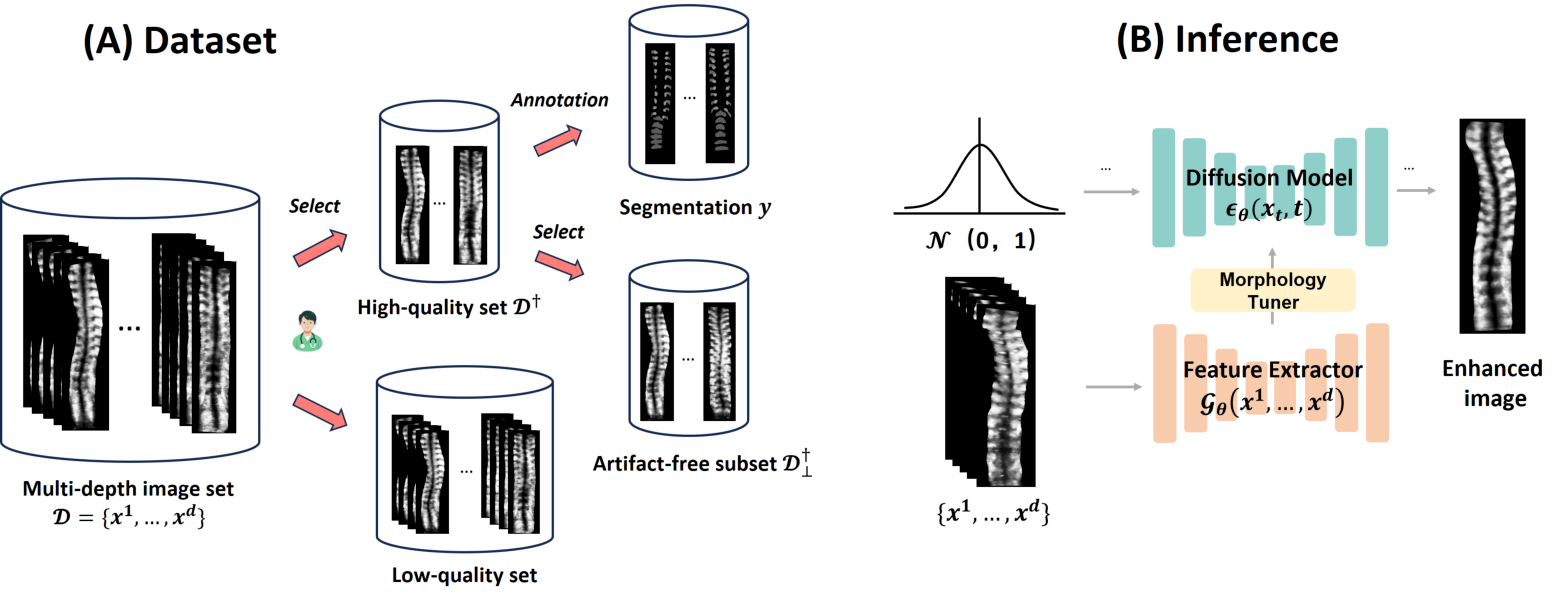}
    \caption{The illustration of the dataset with the objective of learning a distribution of high-quality VPI images based on observations made at different depths. Ultrasound experts select the optimal VPI image and annotate the spine region. A subset of artifact-free images is further selected from the high-quality dataset to facilitate the mapping of the artifact-free image distribution. 
    During inference, the pre-trained diffusion model denoises samples using the prior structural information acquired from the feature extractor through a morphological tuner.
    The resulting images ensure consistency with the original image regarding spinal pose description and the anatomies while providing a visually enhanced perception of the spine.}
    \label{introduction_dataset}
\end{figure*}

\subsection{Image Enhancement in Medical Imaging}

Medical image enhancement aims to improve the visual clarity, sharpness, contrast, and overall quality of medical images, making them more useful for medical professionals in diagnosing and treating various health conditions. Numerous traditional algorithms for image enhancement mainly focus on contrast enhancement or noise filtering \cite{salem2019medical, firoz2016medical}. However, processing the entire image without discrimination consideration makes them fail to preserve detailed structure or suffer from over-smooth issues. 

Due to the superiority of deeply learned features, deep learning frameworks have become increasingly popular, achieving notable performance in medical image enhancement. Supervised learning-based enhancement methods directly learn to map the low-quality images to the corresponding high-quality image. However, these methods suffer from low generalization as collecting paired, low-high medical image datasets is impractical in clinical applications. As an alternative, generative adversarial networks have attracted increasing attention in image enhancement as they do not rely on paired images for training. Based on CycleGAN \cite{zhu2017unpaired}, StillGAN was developed to effectively map from a low-quality domain to a high-quality domain or from a domain with non-uniform illumination to a high-quality one \cite{ma2021structure}. Bobrow et al. \cite{bobrow2019deeplsr} proposed a domain transformation model for laser speckle reduction between coherent and incoherent illumination. Yang et al. \cite{yang2018low} proposed a perceptual loss function for training the GAN-based model to perform low-dose CT image denoising. Liu et al. \cite{liu2023perception} proposed a self-supervised CycleGAN to perform ultrasound super-resolution with perception consistency. However, GAN-based approaches mainly rely on discriminative models to facilitate enhancement learning via adversarial training, which makes this kind of method suffer from model collapse when the distribution of the input images is different from that of the training dataset. Moreover, directly mapping the low-quality image distribution to the high-quality image distribution without any structural prior knowledge will make the models fail to preserve structure details, resulting in suboptimal enhancement performance. These phenomena motivated us to explore a more effective and robust learning framework with the ability to preserve the spinal structure. 


\subsection{Diffusion-based Medical Image Application}
As mentioned, GAN-based methods struggle with details and robust generation. The recently developed diffusion models provide a superior alternative. By incorporating a forward procedure that maps a target distribution to the Gaussian distribution and a backward distribution process that iteratively reverses the forward mapping, diffusion probability models have been successfully applied in various medical image applications. Most of them focus on image domain transformation. Zhou et al. \cite{zhou2023uxdiff} proposed transfering the ultrasound image to X-ray images via a diffusion model, which maps the ultrasound image distribution to the X-ray image distribution. Lyu et al. \cite{lyu2022conversion} examined the performance of the DPM-based model utilizing various sampling strategies for transforming CT and MRI. Özbey et al. \cite{ozbey2023unsupervised} proposed a diffusion model with cycle-consistent architecture for efficient MRI-CT image translation. 

Similar to domain transformation, the diffusion probability model can be directly utilized for image enhancement when defining the high-quality and low-quality datasets as two different domains. Song et al. \cite{song2021scorebased} proposed a model by introducing different measurement processes to the conditional diffusion probability model for CT-based metal artifact removal and sparse-view CT generation. Chung et al. \cite{chung2022score} proposed a diffusion model with denoising score matching for high-quality MRI reconstruction. Stevens et al. \cite{stevens2024dehazing} proposed a double-branch diffusion model for unsupervised haze removal of ultrasound cardiac images. The aforementioned algorithms show the superiority of diffusion models for enhanced high-quality image generation. However, all of them only consider single image mapping when performing high-quality image generation, which means the 3D information in medical data is not explored. In this paper, we argue that an image enhancement framework can greatly benefit from the 3D volume information to further improve the image quality. After considering the computational burden, we proposed a multi-to-one mapping strategy that utilizes various VPI images from different depths of the 3D volume to generate high-quality VPI images. 

\section{problem statement and method}

In this section, we first introduce the composition of the ultrasound dataset. Then, we shift our focus to the training pipeline, providing detailed insights into how we leverage spine segmentation priors and multi-depth images for enhancement and discussing the acquisition of an enhanced image in the real clinical setting.

Throughout this work, we consider a dataset setting denoted as $\mathcal{D} = \{(x^{1}, ..., x^{d})^n\}^N_{n=1} $ where $x^{d}$ represents one of $N$ data point formed by the depth profile $d \in \{ 1, ..., D\} $ (Fig \ref{introduction_dataset} (A)). The optimal images in terms of the best visual quality form the high-quality dataset denoted as $\mathcal{D}^{\dagger} = \{ (x^{\dagger}, y)^n\}^N_{n=1}$ where $y$ defines the ground-truth annotation of spine features.
In $\mathcal{D}^{\dagger }$, we further divide it into an artifact-free subset based on the presence of scanning artifacts, denoted as $\mathcal{D}^{\dagger }_{\bot}$.
The task is to learn a reconstruction network $f_{\theta}(x^1, ..., x^d)$, which recovers the blurred or missing regions presented in original VPI images and further enhances the visibility of vertebral structures. To this end, we treat the VPI image enhancement as a multi-depth information fusion task in which the posterior distribution of enhanced VPI images given multi-depth VPI images are estimated using a diffusion model, enabling multi-to-one mapping. 

\begin{figure*}
    \centering
\includegraphics[width=0.98\textwidth,height=0.4\textwidth]{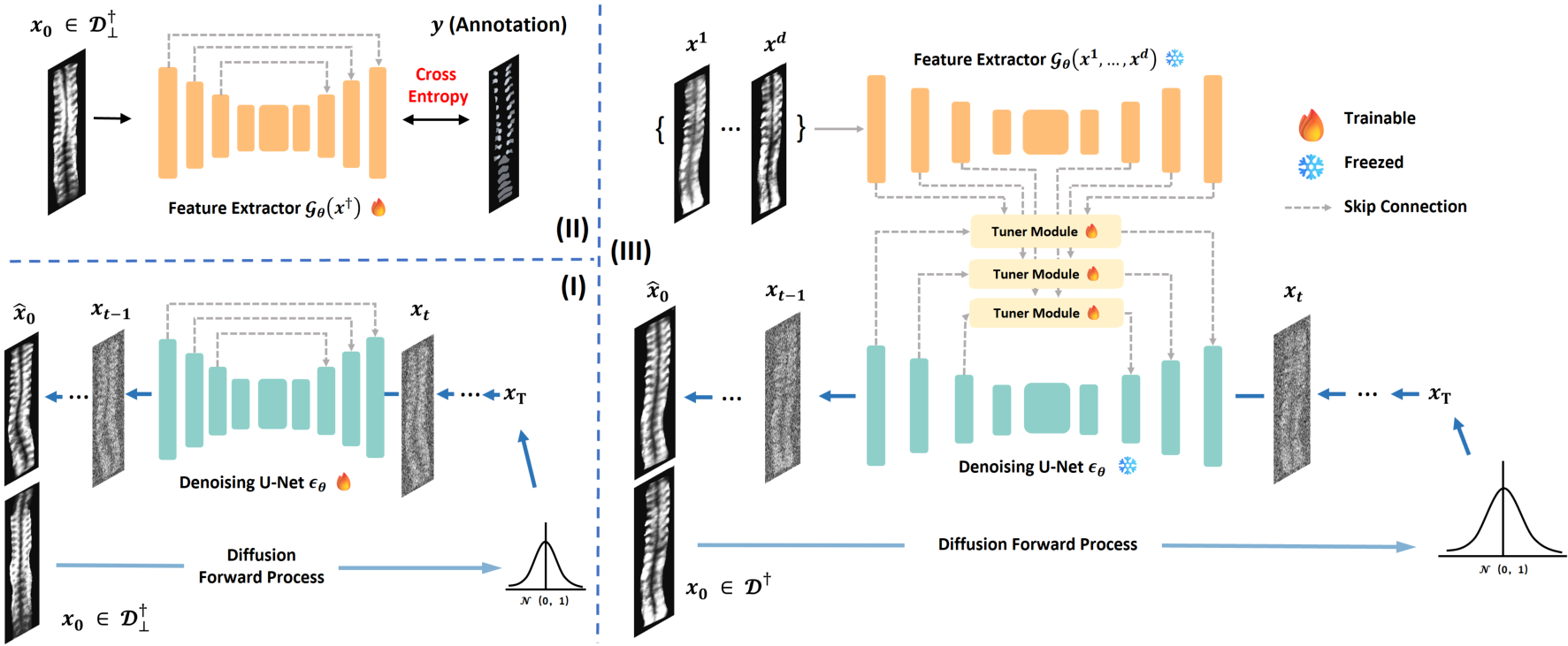}
    \caption{The structure diagram of the proposed enhancement pipeline. At stage I, the diffusion model is trained exclusively on the artifact-free image domain to approximate the high-quality image distribution. At stage II, we train a feature extractor identical to the structure of the diffusion model to capture the accurate morphological information in the ultrasound image. At stage III, with the frozen diffusion model and feature extractor, the high-fidelity structure of the spine can be reconstructed by optimizing the separate tuner module. The tuner receives different levels of latent features extracted from images at different depths and integrates them with the features in the diffusion model to achieve high-quality and high-fidelity image synthesis.}
    \label{method_pipeline}
\end{figure*}

\subsection{High-quality Diffusion Model}
The diffusion model aims to estimate the distribution of VPI images, which are free from artifacts and of high quality. To achieve this, the model begins by gradually adding various levels of Gaussian noise to an artifact-free image $x_{0}^{\dagger } \in \mathcal{D}^{\dagger }_{\bot}$ through a T-step Markov chain \cite{ho2020denoising}. The intermediate variable in step $t$ is formulated as:
\begin{equation}
    x_{t}^{\dagger }=\sqrt{\alpha_{t}}x_{0}^{\dagger}+\sqrt{1-\alpha_{t}} \epsilon^{(t)}
\end{equation}
where $\alpha_{t}$ is a preset constant to control the standard deviation in the current time distribution. $\epsilon^{(t)}$ is the random Gaussian noise added into the current iteration. 
In the diffusion process, a parameterized U-net is widely adopted to model the data distribution $p_{\theta}(x^{\dagger }_{0})$ by maximizing the variational lower-bound:
\begin{equation}
\begin{aligned}
    \mathbb{E}[-\mathrm{log}(p_{\theta}(\mathrm{x}^{\dagger }_{0}))]\le \mathbb{E}_q[-\mathrm{log}(p(\mathrm{x}^{\dagger }_{\mathbf{T}})) 
    - \sum_{t\ge 1}\mathrm{log} 
    \frac{p_{\theta}(\mathrm{x}^{\dagger }_{t-1}|\mathrm{x}^{\dagger }_{t})}{q(\mathrm{x}^{\dagger }_{t}|\mathrm{x}^{\dagger }_{t-1})}  ]
\end{aligned}
\end{equation}
To approach the target distribution, we can estimate the direction of the current time distribution and update the distribution accordingly. This process starts from a random Gaussian noise $x_\mathbf{T}$. By using an iterative sampling approach, we can progressively approximate the target distribution backward. To speed up the sampling, we generalize the original Markovian iterative process as a non-Markovian process \cite{song2020denoising}, which is denoted as:
\begin{equation}
    \hat{x}_{t-1}^{\dagger } = \sqrt{\frac{\alpha_{t-1} }{\alpha_{t}} }\hat{x}_{t}^{\dagger } - (\sqrt{\frac{(1 - \alpha_{t}) \alpha_{t-1}}{\alpha_{t}} } - \sqrt{1 - \alpha_{t-1}})\epsilon_{\theta}^{(t)}(\hat{x}_{t}^{\dagger }) 
\label{DDIM_formulate}
\end{equation}
where $\hat{x}$ indicates the reconstructed sample after the revere process. A pre-trained high-quality VPI diffusion model, trained on a high-quality and artifact-free subset $\mathcal{D}^{\dagger }_{\bot}$, can be treated as the foundation model for downstream morphology editing applications \cite{zhang2023adding}. 
The objective of the foundation model is to synthesize the ultrasound coronal image with the removal of artifacts commonly seen in ultrasound images. 
While the foundation model produces high-quality VPI images, their fidelity may be limited as they fail to preserve the original image's content. This limitation arises from the absence of the spinal pose description and anatomical information of the original image in the generative process.
To achieve precise and consistent VPI image synthesis, it is imperative to meticulously incorporate the spinal pose description and intricate morphology information from the original image while calibrating the generative process.
To this end, we propose combining morphological information from multi-depth images into a unified representation.
After that, the fused representation can serve as a crucial condition for guiding the generation of a high-fidelity VPI image through a controllable tuner module.

\subsection{Multi-depth Fusion Mechanism}

\begin{figure}
    \centering
    \includegraphics[width=0.45\textwidth,height=0.33\textwidth]{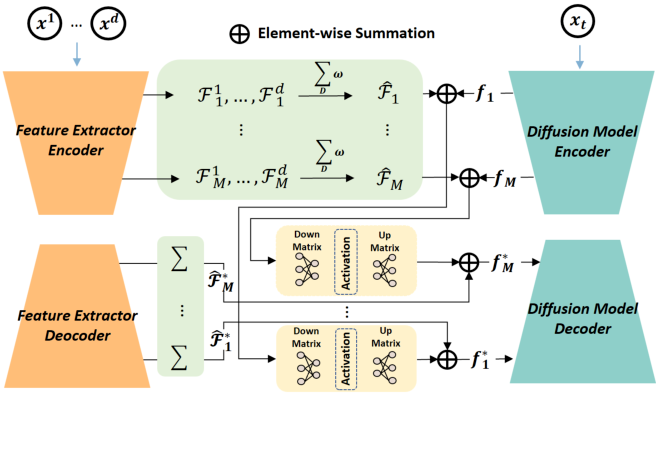}
    \caption{The fusion operation is employed on the latent spaces between the diffusion model and the feature extractor. The corresponding features in the encoder at different levels are passed to the tuner as depth-wise weighted summation. The output of the tuner is further added to the corresponding features of the decoder and then serves as the condition for calibrating the trajectory of the reverse process.}
    \label{method_fusion_pipeline}
\end{figure}

To introduce morphological information into the generative process, we utilize spine segmentation for morphology editing to effectively manipulate the shape, size, and inter-vertebral topology. This involves training an additional feature extractor to capture the variability of morphological features in the latent space. Unlike estimating the distribution of natural images, the feature extractor's objective is to learn regularized latent features that represent shape, structural, and topological information under segmentation supervision. 
We adopt a unified architectural design for the feature extractor to facilitate combining features from the diffusion model and feature extractor, aligning it with the architecture in the high-quality diffusion model. We provide the illustration of the fusion procedure in Fig \ref{method_fusion_pipeline}.
Let $\mathcal{G}_{\theta}(\cdot)$ be the feature extractor with parameters $\theta$, which transforms the multi-depth VPI images into corresponding latent features. We can represent the latent morphology-specific features as follows:
\begin{equation}
 \widehat{y} = \mathcal{G}_{\theta}(x|\mathcal{F}_1, ..., \mathcal{F}_M; \mathcal{F}_M^{\ast }, ..., \mathcal{F}_1^{\ast })
\end{equation}
where $\widehat{y}$ is the predicted spinal segmentation. $\mathcal{F}$ and $\mathcal{F}^{\ast }$ are the intermediate features in the extractor encoder and decoder, respectively. $M$ is the total number of blocks in the encoder. The high-dimensional features at different levels are trained to estimate possibilities around the corresponding anatomical structure's location. However, projecting ultrasound volume data at a single depth may offer a limited perspective on vertebral morphology. A depth that is too shallow may exclude the projection of ribs and transverse processes. In contrast, an excessive depth can result in the loss of intact anatomical structures in the vertebral body.
Instead of directly injecting the multi-depth features into the diffusion model for morphology editing, we propose to fuse multi-depth features into a unified representation first:
\begin{equation}
\widehat{\mathcal{F}}_i^{(\ast )} = \sum_{d=1}^{D}{w_d\mathcal{F}^{(\ast ), d}_i} \quad 1 \le i \le M
\end{equation}
Here, $w$ denotes the weights that balance the strength of latent features from different depths, and $\sum_{d=1}^{D}w_d = 1$. Weighted summation allows the model to prioritize depth images based on their perceived accuracy in the overall anatomical structure. Then, the fusion operation combines the features from each depth image, emphasizing those with higher weights, accurately representing the spinal pose description and the anatomies.

\subsection{Morphology Tuner Module}
Maintaining the consistency of high-frequency information with the enhanced-before image is crucial during the generation process. It refers to the high-quality and high-fidelity image generation. To achieve controllable and high-fidelity generation, tuning paradigms have demonstrated their efficacy in directly editing the conditioning information to adjust the trajectory of the generation process \cite{jiang2024scedit, jiang2024res}. The control conditions consist of fused features extracted from VPI images at various depths, representing the shape of ribs, transverse processes, and lumbar lumps. Moreover, the skip connections have proven beneficial in complementing the high-frequency details within diffusion models \cite{si2024freeu}. 
Inspired by these advancements, we present a morphology tuning module integrated within the skip connections to enable precise control over the generated contours of anatomical structures. Given the outputs of the $i$-th encoder block and decoder block in the diffusion network are $\mathit{f}_{i}$ and $\mathit{f}^{\ast}_{i}$, the original skip connection can be formulated as follows:
\begin{equation}
\mathit{f}^{\ast  }_{i} = Concat([\mathit{f}_{i}, \mathit{f}^{\ast  }_{i}]) \quad 1 \le i \le M
\end{equation}
The learnable tuning module includes a down-sampling matrix $\mathrm{W}_{down}$ that takes the sum of the condition $\widehat{\mathcal{F}}_i$ and the input $x_i$, followed by a non-linear activation function $\phi$ and an up-sampling projection matrix $\mathrm{W}_{up}$. The tuning operation can be formulated as:
\begin{equation}
\mathbf{T} (f_i, \widehat{\mathcal{F}}_i) = \mathrm{W}_{up}\phi(\mathrm{W}_{down}(f_i+\widehat{\mathcal{F}}_i))
\end{equation}
The tuner module first reduces the channel dimension before expanding it back to the original size. During this process, the high-frequency features from the encoder are combined with the fused features from the extractor, which serve as the condition for structure-guided generation. Our tuning strategy is simple but parameter-efficient in introducing morphological information to achieve high-fidelity image generation.
Finally, the skip connection can be reformulated as:
\begin{equation}
\mathit{f}^{\ast  }_{i} = Concat([ \mathbf{T}(\mathit{f}_{i}, \widehat{\mathcal{F}}_i) + \widehat{\mathcal{F}}_i^{\ast }), \mathit{f}^{\ast  }_{i}]) \quad 1 \le i \le M
\end{equation}

\subsection{Training and Sampling}

\subsubsection{Training} There are three stages for training the framework (Fig \ref{method_pipeline}). At the first stage, a U-net-based diffusion network as a foundation model is trained using the artifact-free images in $\mathcal{D}^{\dagger }_{\bot}$. The model architecture is followed in \cite{nichol2021improved}. We minimize the upper bound on the variation of the negative log-likelihood by estimating the $l_2$ loss between the predicted noise and real noise and reconstruction loss between $x^{\dagger}_{0}$ and predicted $\hat{x}^{\dagger }_{0}$:
\begin{equation}
\begin{aligned}
    \mathcal{L} = \mathbb{E}_{\mathcal{D}^{\dagger}, \epsilon, t}\left \|\epsilon -  \epsilon_{\theta}(x^{\dagger }_{t}, t) \right \| ^2_2 \\
    + \mathbb{E}_{\mathcal{D}^{\dagger}, \epsilon, t} \left \| x^{\dagger }_{0} - \frac{x^{\dagger }_{t} -  \sqrt{1-\alpha_{t}} \epsilon_{\theta}(x^{\dagger }_{t}, t)}{ \sqrt{\alpha_{t}}}  \right \| ^2_2  &   
\end{aligned}
\end{equation}
The optimal images with their corresponding segmentation masks are used in the second stage to train a segmentation network. As the initial feature extractor, the segmentation network can capture the contextual information among multi-depth images for editable image generation. The feature extractor has the same components as the foundation model, comprising an encoder and a decoder. The intermediate features in the encoder are fused and then fed to the pre-trained diffusion network. The optimization objective is to use the semantic information in ground-truth labels $y$ to map the multi-depth input to a unified vector representing the ultrasound morphological structure of the spine. To supervise the learning of spine features, the loss function for the feature extractor is based on pixel-wise Cross Entropy (CE) loss:
\begin{equation}
    \mathcal{L}_{\mathcal{G}} = -\sum_{i=1}^{N} y_i \log(\hat{y}_i)
\end{equation}
Consequently, the frozen diffusion model can be fine-tuned using the multi-depth fused features. The fine-tuning is performed on the $\mathcal{D}^{\dagger }$. Despite scanning artifacts in the dataset, the learning of the artifact-free domain will remain unaffected as the backbone network of the diffusion model is frozen. As a result, the model can focus on refining its generation trajectory, producing high-fidelity images whose structure closely aligns with the original, including high-frequency details and overall spine curvature. The parameters in the tuner module are optimized according to the formulation as follows:
\begin{equation}
    \nabla_{\mathbf{T} }\;\mathbb{E}_{\mathcal{D}^{\dagger  }, \epsilon, t}\left \|\epsilon -  \epsilon_{\theta}(x^{\dagger }_t, t, \mathcal{G}_{\theta}(x^1, ... ,x^d) \right \| ^2_2
\end{equation}

\subsubsection{Sampling} 


As presented in Fig \ref{introduction_dataset} (B), image enhancement starts by sampling random Gaussian noise. 
The feature extractor takes ${x^1, ... , x^d}$ as input (including the optimal image), and the intermediate features are combined with the same-level features in the trained diffusion model. The trained model then predicts and removes the added noise at each time step iteratively. In contrast to the original sampling formulation in \cite{song2020denoising}, we require a deterministic sampling process to eliminate unwanted morphological variability throughout the enhancement. As a result, we remove the noise term from the original formulation. This deterministic approach guarantees the structural consistency between the enhanced and original images. 
For each iteration, the encoded morphological information from multi-depth images is injected into the noise predictor via a tuner module. After $T$ times of denoising iteration, the final enhanced image with a high visual perception of vertebral structure is obtained and used to identify the critical landmarks for accurate UCA measurements.

\section{Experiments}

\subsection{Dataset and Experimental Details}
We collected ultrasound spine images using the Scolioscan system. Informed consent was obtained before the scanning session. The dataset consists of 432 3D ultrasound volume data from 191 patients. For each case, we obtained five 2D coronal images at different depths using the volume projection imaging technique. As summarized in Table \ref{dataset}, the dataset is divided into two subsets: 1) The high-quality set, referred to as $\mathcal{D}^{\dagger}$, contains 432 optimal images selected by two ultrasound experts. Ground truth segmentation for these images is also provided by the same experts; 2) The low-quality set consists of 1728 images obtained from the remaining depths. It is important to note that the definition of high-quality images is relative to other depth levels. Although optimal images have the best quality, they may still contain common scanning artifacts. 151 subsets of high-quality images without scanning artifacts are used to train the high-quality diffusion model and feature extractor. 90 extra cases are used to fine-tune the diffusion model for high-fidelity image synthesis. The remaining 181 cases form the test set, which includes images acquired from difficult-to-image subjects to evaluate the model's performance comprehensively.

\begin{table}[t]
    \centering 
    \caption{Datasets and evaluation metrics in the experiments.}  
    \renewcommand\arraystretch{1.2}
    \setlength{\tabcolsep}{1.8pt}
    \begin{tabular}{l  l  l  l }  
        \hline 
        Evaluation & Enhancement & Diagnosis (Reliability of UCA) \\
        \hline 
         Metrics & MS-SSIM, PSNR, CNR, SNR; & ICC, MAD, SEM & \\
        \hline 
        \multirow{3}*{Dataset} & \multicolumn{3}{l}{432 optimal (high-quality) images } \\
        & \multicolumn{3}{l}{432 segmentation of spine} \\
        & \multicolumn{3}{l}{1728 low-quality images} \\
        \hline 
        \multirow{2}*{Training} & \multicolumn{3}{l}{Stage I $\&$ II: 151 artifact-free images with segmentation} \\
        & \multicolumn{3}{l}{Stage III: 241 (151 + 90) image groups} \\
        \hline 
        Test & \multicolumn{3}{l}{181 image groups with five different depth images} \\     
        \hline
    \end{tabular}
    \label{dataset}
\end{table}

All images are standardized to 256 $\times$ 256 resolution and maintain the aspect ratio. Data augmentation includes randomly flipping over images. For the segmentation network, we set the channel of the final convolution layer in the decoder to the 4 to predict the pixel-wise multi-class classification (0:Background; 1:Transverse Process; 2:Rib; 3:Lumbar Lump) and exclude the layer of timestep embedding. Apart from that, the structure of the segmentation model remains consistent with the diffusion model. The number of down-sampling and up-sampling operations is 3 in both the encoder and decoder. We performed statistics of the distribution of the most optimal image index values. Based on the most optimal image index distribution, we set the balanced weights from the superficial to deeper depth to [0.109, 0.205, 0.252, 0.259, 0.174]. 
For the diffusion model, the training epochs are 2000 and 1000 in the pre-trained and fine-tuned stages, respectively. The learning rate is set to 1.0 $\times$ $10^{-3}$. We employ the exponential moving average strategy with a decay factor of 0.999 and set the threshold of clipping the gradient norm of the parameters to 1.0. The noising time step is set to 1000, with a cosine noise schedule to prevent sudden noise level fluctuations. The epoch is set to 200 for the feature extractor, and the learning rate is set to 1.0 $\times$ $10^{-4}$. All networks are trained from scratch and implemented on the PyTorch using two NVIDIA A6000 48G GPUs.

\subsection{Experimental Setups} 

\begin{table*}[htbp]
    \centering  
    \caption{Quantitative comparison of enhancement performance across different depths using MS-SSIM and PSNR (mean $\pm $ standard deviation). The best results on MS-SSIM and PSNR are highlighted in bold.}  
    \renewcommand\arraystretch{1.1}
    \setlength{\tabcolsep}{2pt}
    \begin{tabular}{c c c c c c c | c c c c c c}  
        \hline  
            \multirow{2}*{Method} & \multicolumn{6}{c|}{MS-SSIM $\uparrow$}& \multicolumn{6}{c}{PSNR $\uparrow$} \\
             & d=1 & d=2 & d=3 & d=4 & d=5 & Avg & d=1 & d=2 & d=3 & d=4 & d=5 & Avg \\ 
              \hline
              

            StillGAN & 0.68(0.07) & 0.67(0.08) & 0.66(0.09) & 0.67(0.09) & 0.70(0.09) & 0.68(0.09) & 14.02(0.66) & 14.15(0.61) & 14.07(0.65) & 13.95(0.64) & 14.00(0.60) & 14.04(0.63) \\ 
            
            \cline{2-13}
            

            HQG-Net & 0.63(0.08) & 0.71(0.06) & 0.76(0.06) & 0.75(0.06) & 0.69(0.06) & 0.71(0.06) & 14.35(1.08) & 15.38(0.86) & 15.92(1.03) & 15.58(0.95) & 14.67(0.78) & 15.18(0.94) \\
          
            \cline{2-13}
            

            Arc-Net & 0.66(0.08) & 0.75(0.06) & 0.81(0.06) & 0.79(0.05) & 0.71(0.06) & 0.74(0.06) & 14.56(1.25) & 15.87(1.15) & 16.66(1.46) & 16.12(1.21) & 14.90(0.91) & 15.62(1.20) \\
             
             \cline{2-13}
             

            GFE-Net & 0.69(0.08) & 0.77(0.05) & 0.83(0.05) & 0.80(0.05) & 0.73(0.05) & 0.76(0.05)	& \textbf{15.38}(1.14) & 16.78(0.85) & 17.56(1.06) & 16.92(0.87) & 15.60(0.78) & 16.45(0.94) \\
             
             \cline{2-13}
             

             SCR-Net & 0.68(0.07) & 0.77(0.05) & 0.82(0.05) & 0.80(0.05) & 0.73(0.05) & 0.76(0.05) & 15.24(1.10) & 16.55(0.86) & 17.26(1.10) & 16.72(0.92) & 15.48(0.76) & 16.25(0.95) \\
             
             \cline{2-13}
             

             PCE-Net & 0.68(0.07) & 0.76(0.05) & 0.82(0.04) & 0.79(0.04) & 0.72(0.05) & 0.75(0.05) & 15.14(1.10) & 16.59(0.77) & 17.38(0.93) & 16.79(0.82) & 15.50(0.76) & 16.28(0.88) \\  
             
             \cline{2-13}
             
             
             Ours & \textbf{0.71}(0.08) & \textbf{0.83}(0.05) & \textbf{0.90}(0.03) & \textbf{0.86}(0.03) & \textbf{0.77}(0.06) & \textbf{0.81}(0.05) & 15.30(1.06) & \textbf{17.59}(1.07) & \textbf{19.69}(0.95) & \textbf{18.65}(0.96) & \textbf{16.48}(1.07) & \textbf{17.54}(1.02) \\
            
        \hline

    \end{tabular}
    \label{SSIM and PSNR}

\end{table*}

\begin{table*}[htbp]
    \centering  
    \caption{Quantitative comparison of enhanced images using CNR and SNR (mean $\pm $ standard deviation). The best results on CNR and SNR are highlighted in bold.}  
    \begin{threeparttable}
    \renewcommand\arraystretch{1.0}
    \setlength{\tabcolsep=0.2cm}{
    \begin{tabular}{c c c c c c c c c}  
        \hline  
        Method & Original & StillGAN & HQG-Net & Arc-Net & GFE-Net & SCR-Net & PCE-Net & Ours \\
        \hline  
        CNR (dB) $\uparrow $ 
        & 9.17 $\pm$ 2.54 
        & 7.82 $\pm$ 2.1
        & 6.25 $\pm$ 3.22
        & 9.93 $\pm$ 2.67
        & 9.55 $\pm$ 2.96
        & 9.59 $\pm$ 2.77
        & 9.87 $\pm$ 2.52
        & \textbf{16.87 $\pm$ 0.65} \\
        SNR (dB) $\uparrow$ 
        & 26.78 $\pm$ 1.72 
        & 24.05 $\pm$ 1.91
        & 24.48 $\pm$ 2.13
        & 26.15 $\pm$ 2.08
        & 27.75 $\pm$ 1.83
        & 27.2 $\pm$ 1.76
        & 27.15 $\pm$ 2.00
        & \textbf{33.96 $\pm$ 1.72} \\
        \hline  
       \end{tabular}}    
    \end{threeparttable}
    \label{CNR_SNR}
\end{table*}

\begin{figure*}[htbp]
    \centering
\includegraphics[width=0.95\textwidth,height=0.24\textwidth]{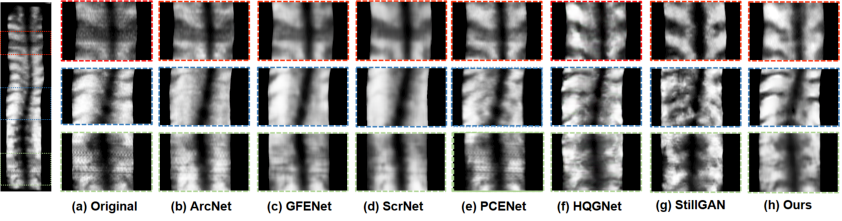}
    \caption{Visualized comparison of image quality enhancement with different methods. The first columns on the left are the optimal ultrasound images selected by the experts. Starting from the second column are the enhancement results of ARC-Net, GFE-Net, SCR-Net, PCE-Net, HQG-Net, StillGAN and ours. The visual perception improvement of the proposed method indicates its versatility in recovering the scanning artifact, eliminating speckle noise, and accentuating the anatomical structure of the spine.}
    \label{result_vis}
\end{figure*}

In this section, we evaluate the proposed enhancement pipeline's performance in terms of image quality and real clinical image diagnosis tasks. We compare the enhancement performance with six baseline algorithms designed for medical image enhancement (StillGAN \cite{ma2021structure}, HQG-Net \cite{he2023hqg}, Arc-Net \cite{li2022annotation}, GFE-Net \cite{li2023generic}, SCR-Net \cite{li2022structure}, and PCE-Net \cite{liu2022degradation}). 
As the literature on VPI image enhancement is not readily available, baseline methods are initially designed for low- and high-fundus image enhancement tasks. The dataset comprises multi-to-one image pairs. To train paired data-based algorithms, the low-quality image is randomly chosen from various depths within the same patient, except the optimal depth. Therefore, there is no need to perform any degradation operation to simulate the real images of low quality for training synthetic image pairs-based methods \cite{li2023generic, liu2022degradation}. 
Furthermore, as noise and artifacts may still be present in optimal images, we calculate no-reference metrics of signal-to-noise (SNR) and contrast-to-noise ratio (CNR) to evaluate the denoising performance:
\begin{equation}    
    SNR = 20\mathrm{log} (\frac{(1 - \mu_{s})}{\sigma_{b}})
\end{equation}
\begin{equation}
    CNR = 20\mathrm{log} (\frac{ \left | \mu_{s} - \mu_{b} \right |}{\sqrt{\sigma_{s}^2 + \sigma_{b}^2} })
\end{equation}
$s$ and $b$ are the signal and background regions of interest. $\mu$ and $\sigma$ are each region's mean and standard deviation. 
The hypoechoic area of the spine (low image intensity) is considered the signal region based on spine segmentation. Therefore, higher SNR and higher CNR are indicative of better image quality.

In the context of ultrasound imaging for scoliosis assessment,
UCA measurement can serve as valuable metrics to quantify
the effectiveness of enhancement algorithms. 
To verify the reliability of UCA, two raters independently perform UCA measurements on the original and enhanced images. Each rater conducts separate measurements for the same set of images. The standardized UCA measurement protocols are strictly followed throughout the measurement process. The statistical analysis includes intra-class correlation coefficients (ICCs), mean absolute difference (MAD), and standard error of measurement (SEM) of measured angles on enhanced and original images.  

\section{Results}

\subsection{Comparison with Enhancement-based Algorithms}

Table \ref{SSIM and PSNR} summarizes the comparison with image enhancement-based algorithms in image similarity across multi-depth ultrasound images. Fig \ref{result_vis} provides the visualized comparison with the algorithms. The original image indicates the most optimal image selected by the experts. HQG-Net and StillGAN are trained using unpaired high-low quality data, which explains why undesired artifacts do not present in the original image are reconstructed in the enhanced image. In contrast, structural-related loss is introduced in ArcNet, GFE-Net, SCR-Net, and PCE-Net to preserve the structural information. However, all of them fail to remove the scanning artifact, which result from the absence of vertebral structure due to the rapid movement of the ultrasound probe. Moreover, ScrNet and GFENet tend to over-smooth the image, making it difficult to recognize the margins of the vertebrae and leading to uncertain line localization deviations. At the same time, StillGAN produces irregular shadows of the vertebra in the reconstructed images.
\begin{table*}[htbp]
    \centering  
    \caption{Inra-rater and Inter-rater variation and reliability of ultrasound curve angle on main thoracic and (throaco)lumbar curves. The ICC values lower than 0.8 and the angle differences greater than 5 degrees are highlighted in bold.}  
    \begin{threeparttable}
    \renewcommand\arraystretch{1.0}
    \setlength{\tabcolsep=0.35cm}{
    \begin{tabular}{c c c c c c c c}  
        \hline  
            \multirow{2}*{UCA Measurement} & \multirow{2}*{Image} & \multicolumn{3}{c}{Main Thoracic Curves}& \multicolumn{3}{c}{(Thoraco) lumbar Curves} \\
            & & ICC$^{a/b}$ (95 $\%$ IC) & MAD($^\circ$) & SEM($^\circ$) & ICC$^{a/b}$ (95$\%$ IC) & MAD($^\circ$) & SEM($^\circ$) \\ 
        \hline
        \multirow{2}*{Intra-rater I} & Original & 0.92 (0.88 - 0.94)$^{a}$ & 3.45 & 2.32 & 0.85 (\textbf{0.78} - 0.89)$^{a}$ & 3.78 & 2.25 \\
        & Enhanced & 0.94 (0.91 - 0.96)$^{a}$ & 3.35 & 2.07 & 0.91 (0.87 – 0.94)$^{a}$ & 3.24 & 1.73 \\
        \multirow{2}*{Intra-rater II} & Original & 0.89 (0.85 – 0.92)$^{a}$ & 4.09 & 2.52 & 0.83 (\textbf{0.76} – 0.88)$^{a}$ & 4.06 & 2.19 \\
        & Enhanced & 0.94 (0.91 – 0.95)$^{a}$ & 3.62 & 1.92 & 0.92 (0.87 – 0.95)$^{a}$ & 3.18 & 1.57 \\
        \hline
            \multirow{2}*{Inter-rater} & Original & 0.85 (\textbf{0.60} – 0.92)$^{b}$ & \textbf{5.19} & 3.67 & 0.82 (\textbf{0.73} – 0.88)$^{b}$ & 3.97 & 2.81 \\
            & Enhanced & 0.91 (0.86 – 0.94)$^{b}$ & 3.80 & 2.69 & 0.89 (0.83 – 0.93)$^{b}$ & 3.34 & 2.36 \\
        \hline
       \end{tabular}}
        \begin{tablenotes}               
             \item ICC$^{a}$: Two-way mixed and absolute agreement; ICC$^{b}$: Two-way random and absolute agreement; \\ IC: Confidence interval; MAD: Mean absolute difference; SEM: Standard error of measurement;   
        \end{tablenotes}    
    \end{threeparttable}
    \label{inter-rater}
\end{table*}
It is worth noting that GFE-Net, PCE-Net, and HQG-Net are initially designed for low-high fundus image enhancement. Their reliance on degraded images for training limits their generality when applied to ultrasound coronal images. 
On the other hand, our proposed pipeline demonstrates outstanding performance in enhancement scenarios. This is represented by an adequate distinction between bilateral transverse processes and their surrounding soft tissue. In an ideal situation, the transverse process's shadow should have a semicircular shape for line placement, while the shadow of the lumbar lump should be the shape of the rectangle. However, due to the customized depth profile, capturing the complete structure of all vertebrae in a single image is difficult. Our feature extractor extracts crucial vertebral information from multi-depth images and integrates them. As a result, we can reconstruct spinal visualizations encompassing the entire spinal column structure.


According to Table \ref{CNR_SNR}, the proposed method also achieves the best quantitative results in terms of SNR and CNR metrics. For CNR, both StillGAN and HQG-Net perform worse than the original images, indicating a lack of sufficient distinction between the spine and surrounding soft tissues. Furthermore, the other baseline methods show minimal improvement ($<$ 1 dB).
As for SNR, the denoising performance of StillGAN, HQG-Net, and Arc-Net is unsatisfactory, suggesting that the background region (high image intensity) contains significant noise that could lead to misidentification of the spinal structure during line placement. In contrast, our method outperforms all other methods, with the highest improvement of $\sim$ 7.18 dB.


\begin{figure}[t]
    \centering
    \includegraphics[width=0.43 \textwidth,height=0.3 \textwidth]{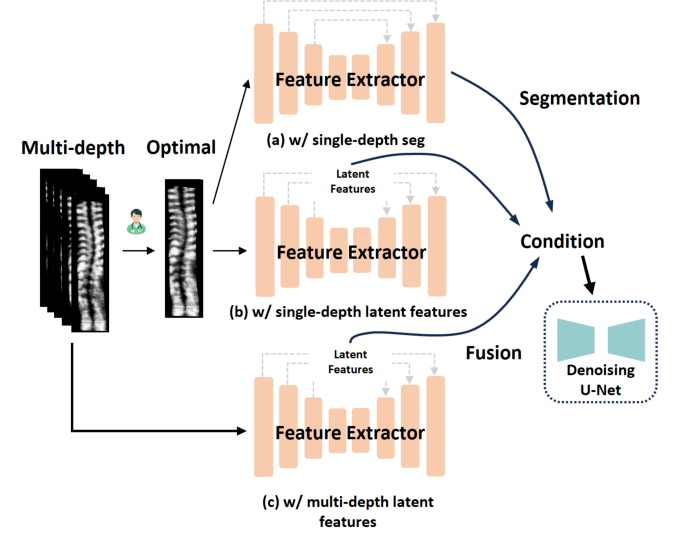}
    \caption{Design of ablation experiment for investigating different condition strategies contributing to image enhancement.}
    \label{ablation_diagram}
\end{figure}
\subsection{Reliability of Ultrasound Curve Angle}
Conveniently deploying UCA measurement on ultrasound coronal images in clinics is a fundamental motivation of this study. 
Table \ref{inter-rater} presents the intra-rater and inter-rater variation and reliability of ultrasound curve angle measurements for the main thoracic and (throaco)lumbar curves.
Regarding intra-rater variation, the measurements performed by Rater I and Rater II demonstrate good consistency between the original and enhanced images. 
However, the confidence intervals for the original images on (thoracic) lumbar are relatively wide, and the low bound is lower than 0.8, indicating poor reliability in the measurements.  
Regarding inter-rater variation, a comparison between the original and enhanced images reveals that the enhanced images yield higher ICC values for both main thoracic and (throaco)lumbar curves. This suggests improved agreement between different raters when using the enhanced images. Moreover, the MAD and SEM values are generally lower in the enhanced image group, indicating reduced differences between raters. It is worth noting that the MAD of the original images on the thoracic curve exceeds 5 degrees. Extra attention is needed when utilizing UCA for scoliosis monitoring in such cases. In contrast, the enhanced image improves the visibility and clarity of the spinal curves, leading to more reliable ($<$ 5 degrees) and consistent angle measurements for scoliosis diagnosis.

\begin{figure}[t]
    \centering
    \includegraphics[width=0.49 \textwidth,height=0.14\textwidth]{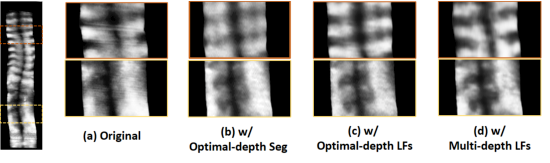}
    \caption{Visualization of the enhancement results for the ablation
study, using the optimal-depth segmentation, optimal-depth latent features, and multi-depth fused features.}
    \label{ablation_fig_vis}
\end{figure}

 \begin{figure}[t]
    \centering
    \includegraphics[width=0.48\textwidth,height=0.1\textwidth]{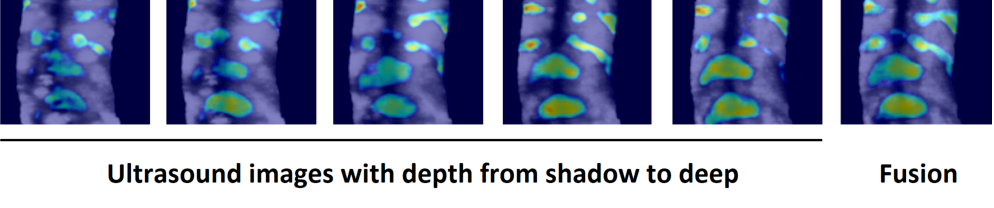}
    \caption{Visualization of the fusion of ultrasound images. The heatmaps are acquired by Grad-CAM \cite{selvaraju2017grad}.}
    \label{ablation_fusion_vis}
\end{figure}

\begin{table}[t]
    \centering  
    \caption{Ablation study of using latent features and fusion strategy.}  
    \begin{threeparttable}
    \renewcommand\arraystretch{1.0}
    \setlength{\tabcolsep=0.2cm}{
    \begin{tabular}{c c c}  
        \hline  
            Original / Conditions & CNR(dB) & SNR(dB) \\ 
        \hline
            Original & 9.17 $\pm$ 2.54 & 26.78 $\pm$ 1.72 \\
            w/ optimal-depth seg & 9.13 $\pm$ 3.66 & 30.92 $\pm$ 3.02 \\
            w/ optimal-depth LFs & 12.29 $\pm$ 1.63 & 30.87 $\pm$ 1.11 \\
            w/ multi-depth LFs & 16.87 $\pm$ 0.65 & 33.96 $\pm$ 1.72 \\
        \hline
       \end{tabular}}   
    \end{threeparttable}
    \label{ablation_table}
\end{table}

\subsection{Ablation Studies}

In this section, we investigate the effectiveness of using optimal-depth latent features (LFs), optimal-depth segmentation, and multi-depth latent features as conditions for image synthesis. Table \ref{ablation_table} summarizes the results of different conditions contributing to image enhancement in terms of CNR and SNR. Fig \ref{ablation_diagram} and \ref{ablation_fig_vis} are the diagrams of the ablation experiment design and the visualization of the enhanced images, respectively. 
For SNR, we initially observe the possibility of mapping the distribution of high-quality images by utilizing prior knowledge of spine segmentation. The reconstructed spine images show improved visual perception compared to the original images. This prompts our investigation into generating high-fidelity images by incorporating spine-specific latent features. We hypothesize that these latent semantic features would provide more reliable indicators of spinal information certainty compared to using segmentation alone. The results substantiate that employing underlying features at different levels outperforms simple segmentation. This approach enables the diffusion models to approximate the target distribution closely. However, as previously mentioned, a single-depth image's semantic representation proves insufficient for reconstructing the entire spine profile due to the limitations of a customized depth profile. Moreover, selecting the optimal image relies heavily on the operator's ultrasound experience and can be time-consuming. To address these limitations, we propose fusing features from images captured at various depths to compensate for missing or incomplete spinal structure information in a single image. 
To elaborate on the ability of the proposed fusion operation, we visualize the multi-depth latent features in the last layer of the decoder using Grad-CAM \cite{selvaraju2017grad}. The results are shown in Fig \ref{ablation_fusion_vis}.
The features in a fused map contain the average of information at different depths and, in turn, can comprehensively represent the overall anatomies of the spine.

\section{Discussion and Conclusion}

In this work, we developed a multi-stage diffusion-based image enhancement pipeline to improve the quality of ultrasound images, explicitly targeting the enhancement of spinal features in scoliosis assessment. Inspired by the supervision with the prior knowledge of morphological information, our proposed model maps the distribution of artifact-free, high-quality images based on a spinal feature extract network with a parameter-efficient tuner module. The clinical significance of our proposed method is multifold, addressing both the reliability and utility of ultrasound-based measurements in a real-world clinical setting.

One of the most critical clinical impacts of the proposed method is its ability to enhance the visibility of spinal features in ultrasound images, significantly reducing intra- and inter-rater variations in measurements. Experts can confidently measure spinal curves, leading to more consistent and reliable assessments. This is crucial for large-scale scoliosis screening, where accuracy and efficiency are paramount. In particular, our enhanced images result in narrower confidence intervals and higher Intraclass Correlation Coefficient (ICC) values for measurement, suggesting improved agreement both within the same rater and between different raters, especially in the thoracic and lumbar regions where original ultrasound images tend to exhibit poor reliability. On the other hand, clinical benefits can be the increased usability of previously challenging or low-quality scans. Many ultrasound images before enhancement are difficult to interpret and often deemed unusable for diagnostic purposes. Our algorithm substantially improves the clarity of these scans, allowing clinicians to extract valuable information from images that were previously considered non-diagnostic. This leads to better utilization of available data, reducing the need for repeat scans and minimizing patient exposure to additional imaging procedures.
Furthermore, the improved image quality also facilitates faster and more efficient measurements, which is especially important in large-scale scoliosis screening programs. By reducing the time required for accurate curve assessment, our method contributes to a more streamlined workflow, enabling the screening of a greater number of patients in a shorter amount of time without sacrificing diagnostic accuracy. This has the potential to significantly impact public health, particularly in regions where large-scale screening is critical for early scoliosis diagnosis and intervention.

In summary, our diffusion-based image enhancement method not only advances the state-of-the-art in medical image processing but also has direct and measurable impacts on clinical practice. By improving the reliability of spinal curve measurements and making difficult-to-image scans usable, our framework offers a robust solution that can improve the accuracy and efficiency of scoliosis diagnosis in clinical settings.

\section*{Conflict of Interest}
Yongping Zheng reports his role as a consultant to Telefield Medical Imaging Limited for the development of Scolioscan, outside the submitted work, and he is also the inventor of a number of patents related to 3D ultrasound imaging for scoliosis, which has been licensed to Telefield Medical Imaging Limited through the Hong Kong Polytechnic University. He is also a director and shareholder of this startup company. All the other authors have no conflicts of interest relevant to this article.

\section*{References}

\bibliographystyle{IEEEtran}
\bibliography{references.bib}

\end{document}